\documentclass[aps,pra,preprint,groupedaddress,showkeys,floatfix]{revtex4}

\begin{document}
\author{Aurelia Cionga$^1$, Fritz Ehlotzky$^{2\thanks{%
Corresponding author: E-mail: Fritz.Ehlotzky@uibk.ac.at}}$, and Gabriela Zloh%
$^1$}
\affiliation{$^1$Institute for Space Sciences, P. O. Box MG-23, \\
R-76900 Bucharest, Romania\\
$^2$Institute for Theoretical Physics, University of Innsbruck,\\
Technikerstrasse 25, A-6020 Innsbruck, Austria}
\title{Electron-atom scattering in a circularly polarized laser field}
\date{November 29, 1999}

\begin{abstract}
We consider electron-atom scattering in a circularly polarized laser field
at sufficiently high electron energies, permitting to describe the
scattering process by the first order Born approximation. Assuming the
radiation field has sufficiently moderate intensities, the laser-dressing of
the hydrogen target atom in its ground state will be treated in second order
perturbation theory. Within this approximation scheme, it is shown that the
nonlinear differential cross sections of free-free transitions do neither
depend on the {\it dynamical phase} $\phi$ of the radiative process nor on
the {\it helicity} of the circularly polarized laser light. Relations to the
corresponding results for linear laser polarization are established. \newline
\end{abstract}
\maketitle
This is the RevTeX shell.

\section{Introduction}

Since the early theoretical work of Bunkin and Fedorov \cite{Bunk} and Kroll
and Watson \cite{Kroll} and the experiments by Weingartshofer et al. \cite
{Wein} a considerable amount of work has been devoted to the investigation
of electron-atom scattering in the presence of a powerful laser field.
Reviews on this topic can be found on the theoretical side in a survey by
one of the present authors \cite{Ehl} and on the experimental situation in a
summary given by Mason \cite{Mason}. Further details can also be found in
the books by Mittleman \cite{Mitt} and by Faisal \cite{Fai}, as well as in
the work by Gavrila \cite{mg}. Initially the atomic target was described by
a static potential but starting with the work of Gersten and Mittleman \cite
{Gerst} the laser dressing of the target was taken into account, treating
the radiation-atom interaction perturbatively. Along the same lines work was
published by Zon \cite{Zon}, Beilin and Zon \cite{Bei} and, in particular,
by Joachain and co-workers in several consecutive papers \cite{Byron}-\cite
{Joa} as well as by Maquet and co-workers \cite{mq1}-\cite{br1}. In all
these investigations a linearly polarized $(LP)$ laser field was considered.
More recently, it became of interest to analyze in some detail the case of a
circularly polarized $(CP)$ laser field \cite{mic}-\cite{pra} to find out,
in particular, whether for $CP$ the nonlinear scattering cross sections
depend explicitly on the {\it dynamical} {\it phase} $\phi $ and the {\it %
helicity} of the radiation field.

It is the purpose of the present work to investigate free-free transitions
on a hydrogen atom for a $CP$ laser field. The scattering process is treated
in the first order Born approximation and the target-dressing by the
radiation field is taken into account in second order perturbation theory.
It will be explicitly shown that in this case the nonlinear differential
cross sections depend neither on the {\it dynamical} {\it phase} $\phi $ nor
on the {\it helicity} of the radiation field. In section II we shall start
our investigations by considering free-free transitions in a CP laser field
on a laser-dressed model potential in order to define the essential
parameters of the process. We shall then investigate in section III in
greater detail and generality the effects of atomic dressing evaluating
first and second order radiative corrections to the bound state. Section IV
will be devoted to a discussion of our numerical results for the angular
distribution and the frequency dependence of the nonlinear signals in
electron-hydrogen scattering in $CP$ laser field. Comparison will be made
between these signals for $CP$ and those for $LP$ fields and the main
differences encountered will be analyzed. The final section will summarize
our findings. Atomic units will be used throughout our investigation.

\section{Scattering on a potential}

We consider free-free transitions for scattering of an electron by the
potential
\begin{equation}
V\left( \vec{r},t\right) =V(r)+\alpha _s\frac{\vec{r}\cdot {\vec{{\cal E}}}%
(t)}{r^3},  \label{pot}
\end{equation}
which may describe a hydrogen atom in a laser field. $V(r)$ denotes the
potential
\begin{equation}
V(r)=-e^{-2r}\left( 1+\frac 1r\right)  \label{SP}
\end{equation}
and $\alpha _s$ is the static polarizability ($\alpha _s$=4.5 a.u. for
hydrogen in its ground state). The second term in eq.(\ref{pot}) describes
approximately the interaction between the electron and the atomic dipole
moment induced by the field. An effective laser-dressed potential of the
form eq.(\ref{pot}) was already used by several authors \cite{Zon}, \cite
{Gelt}, \cite{Varro} for $LP$ fields.

For $CP$ the electric field is given in the dipole approximation by
\begin{equation}
{\vec{{\cal E}}}(t)=i\frac{{\cal E}_0}2\left[ \exp (-i\omega t)\vec{%
\varepsilon}-\exp (i\omega t)\vec{\varepsilon}^{\;*}\right] \equiv \frac{%
{\cal E}_0}{\sqrt{2}}\left( {\vec{e}}_i\sin \omega t-{\vec{e}}_j\cos \omega
t\right)  \label{elf}
\end{equation}
with the polarization vector defined by
\begin{equation}
{\ \vec{\varepsilon}}=\frac{{\vec{e}}_i+i{\vec{e}}_j}{\sqrt{2}},  \label{pol}
\end{equation}
${\cal E}_0$ is the amplitude and $\omega $ the frequency of the electric
field; ${\vec{e}}_i$ and ${\vec{e}}_j$ are unit vectors along two orthogonal
directions in the polarization plane.

In the first order Born approximation, the $S-$matrix element corresponding
to the scattering of the electron on the potential eq.(\ref{pot}) is
\begin{equation}
S_{if}^{B1}=-i\int_{-\infty}^{\infty} \;dt \; <\chi _{\vec{k}_f}(\vec{r}%
,t)|V(\vec{r},t)|\chi _{\vec{k}_i}(\vec{r},t)> .  \label{S-pol}
\end{equation}
$\chi _{\vec{k}_{i,f}}(\vec{r},t)$ are Volkov solutions, which describe the
projectile in the initial and final state, respectively. Since the Volkov
state is written in the velocity gauge, while the electron-dipole
interaction in eq.(\ref{pot}) is written in the length gauge, a gauge factor
would have to be introduced for consistency reasons. In the present
approximation, however, the gauge factors drop out in eq.(\ref{S-pol}).

For an electron of kinetic energy $E_k$ and momentum $\vec{k}$, the Volkov
solution reads
\begin{equation}
\chi _{\vec{k}}(\vec{r},t)=\frac 1{(2\pi )^{3/2}}\exp {\left\{ -iE_kt+i\vec{k%
}\cdot \vec{r}-i\vec{k}\cdot \vec{\alpha}(t)\right\} } ,  \label{vol}
\end{equation}
where $\vec{\alpha}\left( t\right) $ represents the classical oscillation of
the electron in the electric field ${\vec{{\cal E}}}(t)$. In the case of the
above $CP$ laser field this quiver motion is given by
\begin{equation}
{\vec{\alpha}}(t)=\frac{\alpha _0}{\sqrt{2}}\left( {\vec{e}}_i\sin \omega t-{%
\vec{e}}_j\cos \omega t\right) ,
\end{equation}
with $\alpha _0={\cal E}_0/\omega^2$. The Fourier expansion of eq.(\ref{vol}%
) leads to the following series in terms of ordinary Bessel functions, $J_N$%
,
\begin{eqnarray}
\exp \left[ \frac{-i\alpha _0}{\sqrt{2}}{\vec{k}}\cdot \left( {\vec{e}}%
_i\sin \omega t-{\vec{e}}_j\cos \omega t\right) \right] &=&\exp \left\{ -i%
{\cal R}_k\sin \left( \omega t-\phi _k\right) \right\}  \nonumber \\
&=&\sum_NJ_N({\cal R}_k)\exp (-iN\omega t)\exp (iN\phi _k) ,  \label{dev1}
\end{eqnarray}
which is obtained using Graf's addition theorem \cite{Wat}. Accordingly, the
following notations were introduced
\begin{eqnarray}
{\cal R}_k &=&\frac{\alpha _0}{\sqrt{2}}\sqrt{(\vec{k}\cdot {\vec{e}}%
_i)^{\;2}+(\vec{k}\cdot {\vec{e}}_j)^{\;2}}\equiv \alpha _0|\vec{\varepsilon}%
\cdot \vec{k}\;|,  \label{arg} \\
\sin \phi _k &=&\frac{\vec{k}\cdot {\vec{e}}_j}{\sqrt{(\vec{k}\cdot {\vec{e}}%
_i)^{\;2}+(\vec{k}\cdot {\vec{e}}_j)^{\;2}}},  \label{sin} \\
\cos \phi _k &=&\frac{\vec{k}\cdot {\vec{e}}_i}{\sqrt{(\vec{k}\cdot {\vec{e}}%
_i)^{\;2}+(\vec{k}\cdot {\vec{e}}_j)^{\;2}}} .  \label{cos}
\end{eqnarray}
The last two equations lead to
\begin{equation}
\phi _k={\rm arctg}\left( \frac{\vec{k}\cdot {\vec{e}}_j}{\vec{k}\cdot {\vec{%
e}}_i}\right) +l\pi
\end{equation}
with $l$ an integer. We stress that the correct values of $l$ should satisfy
{\bf both} eqs.(\ref{sin}) and (\ref{cos}) in order to be consistent with a
proper use of Graf's addition theorem. By means of the {\it dynamical phase}
$\phi _k$ defined above we get
\begin{equation}
\exp (i\phi_k) = \frac {\vec k \cdot \vec \varepsilon} {|\vec k \cdot \vec %
\varepsilon |}  \label{exp}
\end{equation}
Writing down the expansion eq.(\ref{dev1}) for the momentum transfer of the
scattered electron, $\vec{q}=\vec{k}_i-\vec{k}_f$, one can perform the time
integration in eq.(\ref{S-pol}) to obtain
\begin{equation}
S_{if}^{B1}=\frac i{2\pi }\sum_N\delta \left( E_f-E_i-N\omega \right)
\;f_N^{B1} ,
\end{equation}
where
\begin{equation}
f_N^{B1}=e^{iN\phi _q}\left\{ J_N\left( {\cal R}_q\right)
f_{el}^{B1}(q)-\alpha _s\frac{{\cal E}_0}q\left[ e^{-i\phi _q} \frac{\vec{%
\varepsilon}\cdot \vec{q}}q J_{N-1}\left( {\cal R}_q\right) -e^{i\phi _q}
\frac{\vec{\varepsilon}^{\;*} \cdot \vec{q}}q J_{N+1}\left( {\cal R}%
_q\right) \right] \right\} .  \label{amp}
\end{equation}
$J_N$ denotes a Bessel function of order $N$ and $f_{el}^{B1}$ is the
elastic transition amplitude in the first Born approximation for the static
potential eq.(\ref{SP})
\begin{equation}
f_{el}^{B1}(q)=2\left( q^2+8\right) /\left( q^2+4\right) ^2 ,
\end{equation}
${\cal R}_q$ and $\phi _q$ are defined according to eqs.(\ref{arg}-\ref{cos}%
) using $\vec{q}$ instead of $\vec{k}$.

In the presence of the radiation field the scattered electron may gain or
loose energy equal to $N\omega $, such that $E_f=E_i+N\omega $, where $%
E_{i(f)}$ is the initial (final) energy of the projectile and $N$ is the net
number of photons exchanged (absorbed or emitted) by the colliding system
and the $CP$ field. The energy spectrum of the scattered electrons therefore
consists of the elastic term, corresponding to $N=0$, and of a number of
sidebands, 
each pair of sidebands corresponding to the same value of $|N|$. For
free-free transitions involving $N$ photons one can write down the
differential cross section in terms of the scattering amplitude $f_N^{B1}$
as
\begin{equation}
\frac{d\sigma _N^{CP}}{d\Omega }=\frac{k_f}{k_i}\;|f_N^{B1}|^2 .
\end{equation}
Using eq.(\ref{exp}) 
to rewrite the scattering amplitude eq.(\ref{amp}) one gets by elementary
vector algebra the following form of the differential cross section
\begin{equation}
\frac{d\sigma _N^{CP}}{d\Omega }=\frac{k_f}{k_i}\;\left \vert J_N\left(
{\cal R}_q\right) f_{el}^{B1}(q)-2\alpha _s{\cal E}_0 \frac{|\vec{\varepsilon%
}\cdot \vec{q}|}{q^2} J_N^{\prime }\left( {\cal R}_q\right) \right \vert^2 ,
\label{sec-pot}
\end{equation}
where $J_N^{\prime }$ is the first derivative of the Bessel function with
respect to its argument. It satisfies the relation
\begin{equation}
J_N^{\prime }\left( {\cal R}_q\right) =\frac 12\left[ J_{N-1}\left( {\cal R}%
_q\right) -J_{N+1}\left( {\cal R}_q\right) \right] .
\end{equation}
In this form of the differential cross section it is apparent that the laser
assisted signals in a $CP$ field are neither sensitive to the {\it dynamical}
{\it phase}, $\phi _q$, nor to the {\it helicity} of the photon.

As in the case of linear polarization, the main limitation of the formula
eq.(\ref{sec-pot}) is its failure to describe the well known decreasing of
the target dressing with the increasing of the scattering angle \cite{mq1}.
An improvement, suggested by Milo{\v{s}}evi{\'{c}} {\it et al} \cite{Mil},
consists in replacing the static polarizability, $\alpha _s$, by the
so-called dynamical polarizability
\begin{equation}
\alpha _d=\frac{\alpha _s}{\left( 1+q^2/4\right) ^3} .
\end{equation}
This permits to use eq.(\ref{sec-pot}) at higher scattering angles. Despite
the above limitation, for low frequencies and small scattering angles, eq.(%
\ref{sec-pot}) might be useful as a starting point for the corresponding
investigation of many electron targets for which other methods will likely
be prohibitively difficult to employ.

\section{Electron-atom scattering}

We assume that at moderate laser field intensities, one can describe the
field-atom interaction by time-dependent perturbation theory \cite{Byron}.
We shall use in the following {\it second order perturbation theory} to
describe the hydrogen ground state in the presence of a $CP$ field.
According to Florescu {\it et al.} \cite{fhm}, one can write down an
approximate solution for an electron bound to a Coulomb potential in the
presence of an electromagnetic field as follows
\begin{equation}
|\Psi _1\left( t\right) >=e^{-i{\rm E}_{1}t}\left[ |\psi _{1s}>+|\psi
_{1s}^{(1)}>+|\psi _{1s}^{(2)}>\right] ,  \label{fun}
\end{equation}
where $|\psi _{1s}>$ is the unperturbed ground state of hydrogen, of energy $%
{\rm E}_{1}$, and $|\psi _{1s}^{(1),(2)}>$ denote first and second order
corrections, respectively. According to refs.\cite{fhm} and \cite{f-m} these
corrections can be written in terms of the linear response
\begin{equation}
|{\vec{w}}_{1s}(\Omega )>=-G_C(\Omega )\vec{P}|\psi _{1s}>,  \label{lin}
\end{equation}
and of the quadratic response
\begin{equation}
|w_{ij,1s}(\Omega ^{\prime },\Omega )>=G_C(\Omega ^{\prime })P_iG_C(\Omega
)P_j|\psi _{1s}>.  \label{sec}
\end{equation}
Here $G_C({\Omega })$ is the Coulomb Green's function 
and $\vec{P}$ the momentum operator of the bound electron. For a
monochromatic field there are four values of the argument of the Green
functions necessary in order to write down the approximate solution eq.(\ref
{fun}), namely
\begin{equation}
\Omega ^{\pm }={\rm E}_{1}\pm \omega ,\quad \Omega ^{^{\prime }\pm }={\rm E}%
_{1}\pm 2\omega .  \label{ome}
\end{equation}

On the other hand, as in the case of potential scattering in section II, the
interaction between the $CP$ field and the projectile can be treated exactly
using the Volkov-type solution eq.(\ref{vol}).

We restrict our considerations to high scattering energies where the first
Born approximation in terms of the scattering potential is reliable.
Neglecting exchange effects, we describe this interaction by a static
potential, $V(r,R)$, and the scattering matrix element is then given by
\begin{equation}
S_{if}^{B1}=-i\int_{-\infty }^{+\infty }dt<{\chi }_{{\vec{k}}_f}(t)\Psi
_1(t)|V|{\chi }_{{\vec{k}}_i}(t)\Psi _1(t)>,  \label{GZ}
\end{equation}
where $\Psi _1$ and ${\chi }_{{\vec{k}}_{i,f}}$ are taken from eqs.(\ref{fun}%
) and (\ref{vol}).

The differential cross sections for a process in which $N$ photons are
involved can be written as
\begin{equation}
\frac{d{\sigma }_N^{CP}}{d\Omega }={(2\pi )}^4\frac{k_f{(N)}}{k_i}{|T_N^{CP}|%
}^2,  \label{sed}
\end{equation}
where the transition matrix element, related to the $S$-matrix element eq.(%
\ref{GZ}), has the following general structure
\begin{equation}
T_N^{CP}=\exp \left( iN\phi _q\right) \left[
T_N^{(0)}+T_N^{(1)}+T_N^{(2)}\right] .  \label{tm-cp}
\end{equation}
The first term,
\begin{equation}
T_N^{(0)}=J_N\left( {\cal R}_q\right) <\psi _{1s}|F(\vec{q})|\psi _{1s}> ,
\label{tm0}
\end{equation}
relates to the Bunkin-Fedorov formula \cite{Bunk}, in which the dressing of
the target is neglected. In this case $T_N^{CP}$ reduces to $T_N^{CP}=\exp
\left( iN\phi _q \right) T_N^{(0)}$ and the ordinary Bessel function, $%
J_N\left( {\cal R}_q \right) $, contains all the field dependences of the
transition matrix elements. $F(\vec{q})$ is the form factor operator
\begin{equation}
F(\vec{q})=\frac 1{2\pi ^2q^2}\left[ \exp {(i\vec{q}\cdot \vec{r})}-1\right]
.  \label{ff}
\end{equation}

The other two terms in eq.(\ref{tm-cp}) are due to the dressing of the
atomic state in the $CP$ field and are discussed in the next two subsections.

\subsection{First order dressing of the target}

The second term in eq.(\ref{tm-cp}), $T_N^{(1)}$, is connected to the first
order corrections to the atomic state: {\it one} of the $N$ photons
exchanged between the field and the colliding system interacts with the
bound electron. This photon may be emitted or absorbed and therefore, once
the integration over the coordinates of the projectile was performed, the
general structure of $T_N^{(1)}$ is given by
\begin{equation}
T_N^{(1)}=-\frac{\alpha _0\omega }2\left[ e^{-i\phi _q}\;J_{N-1}({\cal R}%
_q)\;{\cal M}_{at}^{(I)}\left( \Omega ^{+}\right) +e^{i\phi _q}\;J_{N+1}(%
{\cal R}_q)\;{\cal M}_{at}^{(I)}\left( \Omega ^{-}\right) \right] .
\label{tm1}
\end{equation}
The transition matrix element ${\cal M}_{at}^{(I)}\left( \Omega ^{\pm
}\right) $ are related to the exchange of one photon between the atomic
electron and the field. Their expressions read in terms of the linear
response eq.(\ref{lin}) for absorption
\begin{equation}
{\cal M}_{at}^{(I)}\left( \Omega ^{+}\right) = <\psi _{1s}|F(\vec{q})|\vec{%
\varepsilon}\cdot \vec{w}_{1s}\left( \Omega ^{+}\right) >+<\vec{\varepsilon}%
^{\;*}\cdot \vec{w}_{1s}\left( \Omega ^{-}\right) |F(\vec{q})|\psi _{1s}>
\label{def1}
\end{equation}
and emission
\begin{equation}
{\cal M}_{at}^{(I)}\left( \Omega ^{-}\right) = <\psi _{1s}|F(\vec{q})|\vec{%
\varepsilon}^{\;*}\cdot \vec{w}_{1s}\left( \Omega ^{-}\right) >+<\vec{%
\varepsilon}\cdot \vec{w}_{1s}\left( \Omega ^{+}\right) |F(\vec{q})|\psi
_{1s}>,  \label{def2}
\end{equation}
respectively. Using eqs.(8) and (10-12) of ref.\cite{ac1}, one gets
\begin{equation}
{\cal M}_{at}^{(I)}\left( \Omega ^{+}\right) =-\frac{\vec{\varepsilon}\cdot
\vec{q}}{2\pi^2 q^3} {\cal J}_{1,0,1}\left( \tau ^{+},\tau ^{-},q\right),
\hspace*{1cm}{\cal M}_{at}^{(I)}\left( \Omega ^{-}\right) =-\frac{\vec{%
\varepsilon}^{\;*}\cdot \vec{q}}{2\pi^2 q^3} {\cal J}_{1,0,1}\left( \tau
^{-},\tau ^{+},q\right) .  \label{m12}
\end{equation}
The parameters $\tau ^{\pm }$ are related to the parameters $\Omega ^{\pm }$
defined in eq.(\ref{ome}) by
\begin{equation}
\tau ^{\pm }=1/\sqrt{-2\Omega ^{\pm }}.  \label{tau1}
\end{equation}
An analytic expression for ${\cal J}_{1,0,1}$ can be obtained from
eqs.(17)-(22) in Ref.\cite{ac1}. One has
\begin{equation}
{\cal J}_{1,0,1} \left( \tau^{\pm}, \tau^{\mp}, q \right) = {\cal J}%
_{1,0,1}^a \left( q, \tau^{\pm} \right)- {\cal J}_{1,0,1}^b \left( q,
\tau^{\mp} \right) ,  \label{jal}
\end{equation}
where
\begin{eqnarray}
&&{\cal J}_{1,0,1}^a \left( q, \tau \right) = {\cal J}_{1,0,1}^b \left( q,
\tau \right)  \nonumber \\
&& = -\frac{16}{q} \frac{1}{\left(1+\tau\right)^4} \frac{\tau}{2-\tau} Re
\{a^3 F_1 \left( 2-\tau, 1, 3, 3-\tau; \xi, \zeta \right) -\frac{ia^2}{q}
F_1 \left( 2-\tau, 2, 2, 3-\tau; \xi, \zeta \right) \} .
\end{eqnarray}
The foregoing equation is written for frequencies below the ionization
threshold, where $\tau^{\pm}$ are real. $F_1\left(a,b,b^{\prime }, c; x,
y\right)$ is the Appell function of two variables, defined in Ref.\cite{hyp}
and the following notations are used
\begin{eqnarray}
&& a=\frac {2\tau}{1+\tau+iq \tau}, \\
&& \xi=-\frac {1-\tau}{1+\tau}, \quad \zeta=\frac {1-\tau}{1+\tau} \; \frac{
1 -\tau -i q \tau} { 1 +\tau +i q \tau} .
\end{eqnarray}
Our expressions in eq.(\ref{m12}) are equivalent with the ones based on
eq.(18a-c) of Dubois {\it et al} \cite{mq1} and eqs.(11)-(12) of Dubois and
Maquet \cite{mq2}, respectively.

By means of eqs.(\ref{m12}) and (\ref{exp}) one can write down
\begin{equation}
T_N^{(1)}=\frac{\alpha _0\omega }{4\pi ^2q^2}\frac{|\vec{\varepsilon}\cdot
\vec{q}|}q\left[ J_{N-1}({\cal R}_q)\;{\cal J}_{1,0,1}\left( \tau ^{+},\tau
^{-},q\right) +J_{N+1}({\cal R}_q)\;{\cal J}_{1,0,1}\left( \tau ^{-},\tau
^{+},q\right) \right] ,  \label{ta-1}
\end{equation}
which leads to the following transition matrix element
\begin{equation}
T_N^{CP}=\frac{\exp \left( iN\phi _q\right) }{2\pi ^2q^2}\left\{ -\frac{q^2}{%
2} f_{el}^{B1} \left( q \right) J_N\left( {\cal R}_q\right) \; +{\alpha
_0\omega }\frac{|\vec{\varepsilon}\cdot \vec{q}|}q\;J_N^{\prime }\left(
{\cal R}_q\right) \;{\cal J}_{1,0,1}\left( \tau ^{+},\tau ^{-},q\right)
\right\} .  \label{CP}
\end{equation}
To obtain the last expression, we used the following identity
\begin{equation}
{\cal J}_{1,0,1}\left( \tau ^{+},\tau ^{-},q\right) =-{\cal J}_{1,0,1}\left(
\tau ^{-},\tau ^{+},q\right) .
\end{equation}

In this framework, where the first order radiation correction to the ground
state is taken into account only, the differential cross section for a
process in which $N$ photons are exchanged between the colliding system and
the $CP$ laser field is given by
\begin{equation}
\frac{d\sigma _N^{CP}}{d\Omega }=\frac{k_f}{k_i}\; \left\vert f_{el}^{B1}
\left( q \right) J_N\left( {\cal R}_q\right) -2\alpha_0 \omega \frac{|\vec{%
\varepsilon}\cdot \vec{q}|}{q^3}J_N^{\prime }\left( {\cal R}_q\right) {\cal J%
}_{1,0,1}\left( \tau ^{+},\tau ^{-},q\right) \right\vert^2 .  \label{CS}
\end{equation}
We conclude from this result: $(i)$ the {\it dynamical phase} $\phi _q$
drops out and hence has no effect on the differential cross section and $%
(ii) $ due to the appearance of the modulus $|\vec{\varepsilon}\cdot \vec{q}%
| $, the {\it helicity} of the photon is not a relevant parameter.

For the purpose of future reference, we write also down the corresponding
differential cross section for linear polarization
\begin{equation}
\frac{d\sigma _N^{LP}}{d\Omega }=\frac{k_f}{k_i}\; \left\vert f_{el}^{B1}
\left( q \right) J_N\left( \vec{\alpha}_0\cdot \vec{q}\right) -2\alpha
_0\omega \frac{\vec{e}\cdot \vec{q}}{q^3}J_N^{\prime }\left( \vec{\alpha}%
_0\cdot \vec{q}\right) {\cal J}_{1,0,1}\left( \tau ^{+},\tau ^{-},q\right)
\right\vert^2 .  \label{LP}
\end{equation}
To avoid possible confusions, the linear polarization vector was denoted by $%
\vec{e}$. One can see that, apart from the arguments of the Bessel functions
(${\cal R}_q$ instead of $\vec{\alpha}_0\cdot \vec{q}$), the only difference
between eqs. (\ref{CS}) and (\ref{LP}) concerns the angular parts ($|\vec{%
\varepsilon}\cdot \vec{q}|$ instead of $\vec{e}\cdot \vec{q}$).

Based on the low frequency limit in eq.(\ref{jal}), we also mention that the
transition matrix element (\ref{CP}) leads to the expression
\begin{equation}
\frac{d\sigma _N^{CP}}{d\Omega }\simeq \frac{k_f}{k_i}\; \left\vert
f_{el}^{B1}(q) J_N\left( {\cal R}_q\right) - \frac{192}{\left( q^2+4\right)
^3}\left( 1+\frac 8{q^2+4}\right) {\cal E}_0\frac{|\vec{\varepsilon}\cdot
\vec{q}|}{q^2}J_N^{\prime }\left( {\cal R}_q\right) \right\vert^2 ,
\end{equation}
which can be immediately compared with eq.(2.31a) of Byron {\it et al.} \cite
{byr}, evaluated for linear polarization. In addition, for small scattering
angles, where $q \ll 1$, the quantity in front of ${\cal E}_0$ may be
approximated by 9$\simeq 2\alpha _s$. This shows that one may consider eq.(%
\ref{sec-pot}) as the low frequency limit of the differential cross section
eq.(\ref{CS}), valid at small scattering angles.

Our equations (\ref{CS}) and (\ref{LP}) emphasize the importance of the
geometrical relation between the momentum transfer of the scattering
electron and the polarization vector. To make the discussion clear, we shall
choose the quantization axis, $Oz$, along the direction of the initial
momentum of the projectile and the axis $Oy$ in the scattering plane.

It is worthwhile to point out here the correspondences between the three
most frequently considered scattering geometries for $LP$ laser light, namely

\begin{enumerate}
\item  {\bf LP1}: $\vec{e}$ parallel to the initial momentum, ${\vec{e}}%
\;||\;Oz$,

\item  {\bf LP2}: $\vec{e}$ orthogonal to the initial momentum but in the
scattering plane, ${\vec{e}}\;||\;Oy$,

\item  {\bf LP3}: $\vec{e}$ parallel to the momentum transfer, $\vec{e}\;||\;%
\vec{q}$,
\end{enumerate}

and the following configurations involving $CP$

\begin{enumerate}
\item  {\bf CP1}: $\vec{\varepsilon}=(\vec{e}_z+i\vec{e}_x)/\sqrt{2}$, when
the laser beam is propagating in the scattering plane,

\item  {\bf CP2}: $\vec{\varepsilon}=(\vec{e}_x+i\vec{e}_y)/\sqrt{2}$, when
the laser beam is parallel to the direction of the initial momentum, ${\vec{k%
}}_i$,

\item  {\bf CP3}: $\vec{\varepsilon}=(\vec{e}_y+i\vec{e}_z)/\sqrt{2}$ lies
in the scattering plane ($yOz$) and the laser beam propagates on the $Ox$
direction.
\end{enumerate}

We mention that for {\bf CP1} and {\bf CP2} there is only one component of
the $CP$ vector in the scattering plane, while for {\bf CP3} both components
are active. One can immediately see that the following relation holds
\begin{equation}
\left\vert\vec{\varepsilon}_j\cdot \vec{q}|^2=|\vec{e}_j\cdot \vec{q}%
\right\vert^2/2 ,  \label{geo}
\end{equation}
where the index $j$ refers to the above enumerations.

Once we have established these correspondences, we can make a couple of
remarks concerning the relations between laser assisted signals in $CP$ and $%
LP$ laser fields. We shall always refer to $LP$ and $CP$ which are connected
to each other by eq.(\ref{geo}).

The first remark concerns the difference between the laser assisted signals
in $LP$ and $CP$ in the case of the elastic term ($N$=0). For low
frequencies, in the forward direction, the laser assisted signal is smaller
in $CP$ than in $LP$ and the difference is given by
\begin{equation}
\frac{d\sigma _0^{LP}}{d\Omega }-\frac{d\sigma _0^{CP}}{d\Omega }\simeq
\alpha _s \frac {{\cal E}_0^2}{\omega ^2} \frac{\left\vert{\vec e}_j \cdot
\vec q \right\vert^2}{q^2}  \label{dif1}
\end{equation}
in any of the three related configurations.

The second remark is more general and it is valid for any photon frequency.
For weak laser fields at any scattering angles and for moderate laser
intensities at small scattering angles, i.e. whenever the arguments of the
Bessel functions are small, the following relation exists for $|N|\geq 1$
\begin{equation}
\frac{d\sigma _N^{CP}}{d\Omega }\simeq \frac 1{2^{|N|}}\frac{d\sigma _N^{LP}%
}{d\Omega }.  \label{dif2}
\end{equation}

This relation is of particular interest since for $N=\pm 1$ one can recover
in this way the perturbative limit given by
\begin{equation}
\frac{d\sigma _{\pm 1}^{CP}}{d\Omega }=\alpha _0^2\frac{k_f}{k_i}\;\frac{|%
\vec{q}\cdot \vec{\varepsilon}|^2}4\;\left| f_{el}^{B1}\left( q\right) \mp
\frac{2\omega }{q^3}{\cal J}_{1,0,1}\left( \tau ^{+},\tau ^{-},q\right)
\right| ^2.  \label{per}
\end{equation}
The same expression can be obtained by using eqs.(14-16) of Ref.\cite{ac1}.
One should keep in mind that in that paper the photon energy was expressed
in Rydbergs and that eq.(7) of the same paper (devoted to excitation
processes) should be modified for free-free transitions by putting $f_{el}=%
{\cal J}_{10}(q)-1=-q^2f_{el}^{B1}/2$. In the weak field limit we find out
that, on account of the relations eq.(\ref{geo}), the laser assisted signal
involving one $CP$ photon (absorbed/emitted) will be always one half of the
corresponding signal for $LP$. For higher intensities, the deviations from
this relation appear as a signature of nonlinear dynamics.

\subsection{Second order corrections to the ground state}

We shall show in this section that by adding second or higher order terms in
the expansion (\ref{fun}) we get no change in our main conclusion that
neither the dynamical phase $\phi_q$ nor the photon helicity are relevant
parameters in free-free transitions at high scattering energies.

If the second order correction, $\mid \psi_{1s}^{(2)}>$ in eq.(\ref{fun}),
is added to the wave function that describes the ground state of hydrogen in
the laser field, we get the third contribution to the transition matrix
element in eq.(\ref{tm-cp}). After integration over the coordinates of the
projectile, this contribution reads
\begin{eqnarray}
&&T_N^{(2)}=\frac{\alpha _0^2\omega ^2}4\left\{ e^{-2i\phi _q}J_{N-2}\left(
{\cal R}_q\right) {\cal M}_{at}^{(II)}\left( \Omega ^{\prime +},\Omega
^{+}\right) +e^{2i\phi _q}J_{N+2}\left( {\cal R}_q\right) {\cal M}%
_{at}^{(II)}\left( \Omega ^{\prime -},\Omega ^{-}\right) \right.  \nonumber
\\
&& \hspace*{2cm} \left. + J_N \left( {\cal R}_q\right) \left[ \widetilde{%
{\cal M}}_{at}^{(II)}\left( {\rm E_1},\Omega^{-}\right) +\widetilde {{\cal M}%
}_{at}^{(II)}\left( {\rm E_1},\Omega^{+}\right) \right] \right\}.  \label{t2}
\end{eqnarray}
In this expression {\it two} of the $N$ photons exchanged between the field
and the colliding system interact with the bound electron. ${\cal M}%
_{at}^{(II)} \left( \Omega ^{\prime \pm },\Omega ^{\pm }\right)$ are related
to the absorption (upper signs) or emission (lower signs) of {\it both}
photons. Written in terms of the quantities given by eqs.(\ref{lin})-(\ref
{sec}) ${\cal M}_{at}^{(II)}$ have the form
\begin{eqnarray}
&&{\cal M}_{at}^{(II)}\left( \Omega ^{\prime \pm },\Omega ^{\pm }\right)
=\sum_{j,l=1}^3\varepsilon _j\varepsilon _l\left[ <\psi _{1s}|F(\vec{q}%
)|w_{lj,1s}\left( \Omega ^{\prime \pm },\Omega ^{\pm }\right) >\right.
\nonumber \\
&&\hspace*{1.7cm}\left. +<w_{j,1s}(\Omega ^{\mp })|F(\vec{q})|{w}%
_{l,1s}(\Omega ^{\pm })>+<w_{lj,1s}\left( \Omega ^{\prime \mp },\Omega ^{\mp
}\right) |F(\vec{q})|\psi _{1s}>\right] .  \label{defm}
\end{eqnarray}
We stress that the complex conjugate of the polarization vector ${\vec{%
\varepsilon}}$ must be taken in eq.(\ref{defm}) when ${\cal M}%
_{at}^{(II)}\left( \Omega ^{\prime -},\Omega ^{-}\right) $, related to
emission, is computed for a $CP$ laser field. For this polarization ${\vec{%
\varepsilon}}^{\;2}=0$ and the angular behavior of ${\cal M}_{at}^{(II)}$ is
determined by
\begin{equation}
{\cal M}_{at}^{(II)}= \frac{(\vec{\varepsilon}\cdot \vec{q})^2}{2\pi ^2q^4}
{\cal T}_1 \left(\tau^{\prime +},\tau^{\prime -};\tau^+,\tau^-, q \right) ,
\label{taf}
\end{equation}
where ${\cal T}_1$ depends not only on $q$ and $\tau^{\pm}$ but also on
\begin{equation}
\tau^{\prime \pm}=1/\sqrt{-2\Omega^{\prime \;\pm}}  \label{tau2}
\end{equation}
and can be expressed in terms of series of hypergeometric functions, as
shown in the Appendix.

The other two atomic matrix elements in eq.(\ref{t2}), $\widetilde {{\cal M}}%
_{at}^{(II)}$, are related to the processes in which one photon is absorbed
and the other is emitted. They can be constructed by using eq.(\ref{defm})
with the tensor $\widetilde w_{lj}$ (instead of the tensor $w_{lj}$), which
is also defined in Ref.\cite{fhm}. Their angular behavior is different,
namely
\begin{equation}
\widetilde{{\cal M}}_{at}^{(II)}=\frac 1{2\pi ^2q^2}\left[ \frac{|\vec{%
\varepsilon} \cdot \vec{q}\;|^2}{q^2}{\widetilde {{\cal T}}}_1 + |\vec{%
\varepsilon}\;|^2 {\widetilde{{\cal T}}}_2\right] .
\end{equation}
We point out that the radial integrals ${\widetilde {{\cal T}}}_1$ and ${%
\widetilde{{\cal T}}}_2$ must be computed for 
$\Omega^{\prime}={\rm E}_1$. A special work \cite{n0} will be devoted to
their analytic evaluation since a number of technical difficulties that are
related to their singular behavior must be discussed in detail.

Similar to the case of the first order correction in section III.A, one can
express $T_N^{(2)}$ as 
\begin{equation}
T_N^{(2)}=\frac{\alpha _0^2\omega ^2}{8\pi ^2q^2}\left\{ \frac{|\vec{%
\varepsilon}\cdot \vec{q}|^2}{q^2}\; \left[ {\cal T}_1 \;
\left(J_{N+2}\left( {\cal R}_q\right) + J_{N-2}\left( {\cal R}_q\right)
\right) +\widetilde {{\cal T}}_1 \; J_N\left( {\cal R}_q\right) \right] +%
\widetilde {{\cal T}}_2 \; J_N\left( {\cal R}_q\right) \right\} .
\end{equation}
Then, on account of the structure of the transition matrix element including
second order laser dressing of the target, 
\begin{eqnarray}
&&T_N^{CP} =-\frac 1{4\pi ^2}\left\{ f_{el}^{B1}J_N(R_q) -2{\alpha _0\omega }
\frac{|\vec{\varepsilon}\cdot \vec q|}{q^3}\; {\cal J}_{1,0,1}J_N^{\prime }
(R_q) \right.  \nonumber \\
&&\hspace*{1.2cm} -{\alpha _0^2\omega ^2} \frac {|\vec{\varepsilon}\cdot
\vec{q}\;|^2}{2q^4} \left[{\cal T}_1 \; \left(J_{N+2}\left( {\cal R}%
_q\right) + J_{N-2}\left( {\cal R}_q\right) \right) +\widetilde {{\cal T}}_1
\; J_N\left( {\cal R}_q\right) \right] -\left. \frac {\alpha _0^2\omega ^2}{%
2q^2} \widetilde {{\cal T}}_2 \; J_N\left( {\cal R}_q\right) \right\} ,
\end{eqnarray}
one can immediately say that the differential cross section is again
helicity independent and the dynamical phase is not a relevant parameter.

Moreover, based on angular momentum algebra considerations, one can argue
that any contribution to the transition matrix element due to the $j^{{\rm th%
}}$ order perturbative corrections to the atomic state will only contain
terms proportional to $|\vec \varepsilon \cdot \vec q\;|^p \; |\vec %
\varepsilon\;|^{2s}$, where $p$ and $s$ are positive integers such that $%
p+2s=j$. Therefore, as long as the scattering is treated in first order Born
approximation, the helicity will remain an unobservable parameter.

It is interesting to note that the weak field limit of the differential
cross section for two $CP$ photon absorption/emission has a simple angular
dependence given by $|\vec{q}\cdot \vec{\varepsilon}|^4$. Indeed, one finds
\begin{equation}
\frac{d\sigma _{\pm 2}^{CP}}{d\Omega }=\alpha _0^4\frac{k_f}{k_i}\;\frac{|%
\vec{q}\cdot \vec{\varepsilon}|^4}{2^6}\;\left| f_{el}^{B1}-\frac{4\omega }{%
q^3}{\cal J}_{1,0,1}-\frac{4\omega ^2}{q^4}{\cal T}_1\right| ^2.
\label{pers}
\end{equation}

On the contrary, for $LP$ fields where ${\vec{e}}^{\;2}=1$, the atomic
matrix element ${\cal M}_{at}^{(II)}$ has a different angular behavior given
by
\begin{equation}
{\cal M}_{at}^{(II)}=\frac 1{2\pi ^2q^2}\left[ \frac{(\vec{e}\cdot \vec{q})^2%
}{q^2}{\cal T}_1 + {\cal T}_2\right] .
\end{equation}
The amplitudes ${\cal T}_1$ and ${\cal T}_2$ depend on the momentum transfer
of the scattered electron and on the four parameters $\tau^{\prime \pm}$ and
$\tau^{\pm}$, given in eqs.(\ref{tau1}) and (\ref{tau2}). Finally, in the
weak field domain, this leads to the following expressions for the
differential cross sections for two photon absorption/emission in LP fields
\begin{equation}
\frac{d\sigma _{\pm 2}^{LP}}{d\Omega }=\alpha _0^4\frac{k_f}{k_i}\; \frac{1}{%
2^6} \left\vert\left(\vec{q}\cdot \vec e \right)^2\; \left(f_{el}^{B1} -%
\frac{4 \omega}{q^3} {\cal J}_{1,0,1} -\frac{4 \omega^2}{q^4} {\cal T}_1
\right) -\frac{4 \omega^2}{q^2}{\cal T}_2 \right\vert^2 .  \label{perl}
\end{equation}

\section{Results and discussion}

In this section we illustrate our results by considering the numerical
evaluation of the nonlinear differential cross sections for the elastic term
($N=0$) and the next two sidebands ($N=1$ and $N=2$). We focus our
discussion on the geometries denoted earlier by {\bf CP3} and {\bf LP3}
because in these geometries the coupling between the laser field and the
colliding system is particularly strong.

The angular distributions of the scattered electrons with final energies
given by $E_f=E_i+N\omega$ are shown in Figure 1 for the three values of $N$%
: 0, 1, and 2. We have chosen two frequencies in the optical domain, namely $%
\omega$=5 eV and $\omega$=2~eV and our results are evaluated for the laser
field strength ${\cal E}_0$=10$^8$ V/cm and the initial scattering energy $%
E_i$=100 eV.

In the left hand panels $a_1$, $b_1$, and $c_1$ of Figure 1 the laser
frequency is $\omega$=5 eV and the quiver amplitude, $\alpha_0$, takes the
value 0.58 a.u., corresponding to the perturbative regime. In panel $a_1$,
at small scattering angles where the dressing of the target is important,
the differential cross section $d\sigma_0/d\Omega$ exceeds the field-free
signal for both linear and circular polarizations. The nonlinear signals, $%
d\sigma_1/d\Omega$, belonging to the final energy $E_f$=105 eV are presented
in panel $b_1$. Here we recognize that for small arguments of the Bessel
functions the assisted signals for $CP$ have half the value of the signals
for $LP$. Finally, in panel $c_1$ we find large differences between the $CP$
and $LP$ signals. To understand this different behavior we focus on the
dominant contributions to the differential cross sections, given by (\ref
{pers}) and (\ref{perl}), respectively. Moreover, we note that only ${\cal T}%
_2$ of eq.(\ref{perl}) has a pole at $\tau^{\prime}$=2 while it does not
appear in eq.(\ref{pers}). Hence it becomes clear that the enhancement of
the $LP$ signal originates in a two photon virtual transition between the
ground state and the first excited state (E$_2$-E$_1$=10.2 eV).
Considerations of angular momentum algebra can be used to show that such
transitions are forbidden if the two photons have circular polarization.

In Figure 1 (panels $a_2$, $b_2$, and $c_2$) we also present the angular
distributions for the second laser frequency, $\omega$=2 eV. In this case
the amplitude of the quiver motion takes the value $\alpha_0$=3.6 a.u. and
the nonlinear dynamical behavior becomes apparent. Therefore the angular
distributions are considerably different from those of the previous case. We
point out that our formula (\ref{dif1}) reproduces quite well the
differences between $LP$ and $CP$ signals for $N$=0, since here second order
corrections are of minor importance. For $N$=1 the $CP$ signals are again
one half of the $LP$ ones at small scattering angles, as is shown in the
window inserted in panel $b_1$. With increasing scattering angle, the
argument of the Bessel functions increases and nonlinear contributions
become important. The present frequency, $\omega$=2 eV, is too small for
establishing a two-photon resonance and the chosen intensity is not strong
enough for higher order contributions. Therefore $LP$ and $CP$ signals
remain comparable for $N$=2. We think that the differences between our data
for $\omega$=2 eV and those published earlier for the same parameters \cite
{cp1} are due to spurious phase effects present in the calculations of that
work.

We shall next discuss the resonance structure of the first ($N$=1) and of
the second ($N$=2) sidebands considering a sufficiently small scattering
angle, $\theta$=5$^\circ$, such that the target dressing effects are
relevant.

In Figure 2 the resonance structures of $d\sigma_1/d\Omega$ are shown for
one photon absorption in the geometry {\bf CP3}. We restrict ourselves to
the weak field domain and we normalize the signals with respect to the
intensity of the laser field. The differential cross sections exhibit a
number of resonance peaks corresponding to $\omega=|{\rm E}_{1}|(1-n^{-2})$,
where $|{\rm E}_{1}|$ is the binding energy of the ground state and $n$ is
the principal quantum number. They correspond to the poles in the analytic
expression of ${\cal J}_{1,0,1}$ in eq.(\ref{jal}). At very low frequencies
the major contribution stems from the Bunkin-Fedorov term (dotted curve).
The first minimum of the differential cross sections, close to $\omega$=2
eV, comes from an interference between the atomic and the electronic term;
the other minima, located between two consecutive resonances, are related to
the contribution of the first order dressing correction to the ground state
in eq.(\ref{fun}). Since the relation, eq.(\ref{dif2}), holds in the
perturbative domain, the resonance structures for the related scattering
geometry, {\bf LP3}, are obtained by a vertical upshift of these curves by a
factor $\log_{10} (2)$.

The frequency dependences of the next sideband ($N$=2) exhibit two series of
resonances. One photon resonances are located above 10.2 eV as discussed
earlier. In addition, a second series of resonances, located between 5.1 and
6.8 eV, is predicted. It corresponds to {\it two} photon virtual transitions
to excited states. This further series of resonances is related to second
order corrections to the ground state (\ref{fun}) and is presented in our
Figure 3. The panel $a$ refers to the geometry {\bf CP3}, while the other
one to the geometry {\bf LP3}. As discussed before, the resonance located at
$\omega$=5.1 eV is present for $LP$ only. One can explicitly see in our
figures that the dressing effects are increasing for increasing frequencies
of the laser beam.

\section{Summary and conclusions}

In the present work we have investigated scattering of electrons by hydrogen
atoms in the presence of a circularly polarized laser field. For comparison,
we also considered linearly polarized laser light. Since we assumed the
scattered electrons to have initially some $100$ eV kinetic energy, we were
permitted to treat the scattering process in first order Born approximation.
The laser dressing of the atomic target was treated in second order
perturbation theory, while that of the scattering electron was described by
a Volkov solution. Within this approximation scheme, we were able to show
that the nonlinear cross sections $d\sigma _N^{CP}/d\Omega $ neither depend
on the dynamical phase $\phi_q $, contrary to what was predicted by earlier
work on this topic \cite{cp1},\cite{cp3}, nor is there any indication of
circular dichroism. In our derivation of the above findings we devoted
particular attention to the proper definition of the phases in Graf's
addition theorem of Bessel functions, basing our considerations on the
corresponding definitions in Watson's book \cite{Wat}. This was outlined, in
particular, in section II of this work. As we found out, it is very crucial
to make a careful analysis of the phase relations in the above treatment for
otherwise quite easily spurious phase dpendences can creep in to finally
simulate circular dichroism in the process studied above. Besides, we took
advantage of our analysis to also make a comparison between nonlinear
electron-atom scattering in a circularly and a linearly polarized laser beam
of equal frequency and intensity. Among other differences between these two
cases, we were able to show that for weak fields, at any scattering angles,
and for moderate fields, at small scattering angles, $d\sigma
_N^{CP}/d\Omega $ are always smaller than $d\sigma _N^{LP}/d\Omega$.
Moreover, the resonance structures of the two cross sections are different,
in particular, there are more resonances in the linear than in the circular
case. Although one can qualitatively understand these differences by using
angular momentum considerations, we have explicitly shown in the Appendix
how the additional resonance in the case of linear polarization comes about.

Nevertheless, we should stress that a possible phase dependence may occur if
the scattering process is treated beyond the first order Born approximation
\cite{Mitt}. In this case the appearance of imaginary parts of the
scattering amplitude may lead to phase dependences and eventually circular
dichroism in the scattering of electrons by atoms in circularly polarized
laser light. In a forthcoming paper we shall show that, for a particular
laser configuration, circular dichroism due to the target dressing can be
predicted for high scattering energies, choosing the laser frequency and the
scattering geometry in an appropriate way.

\section{Appendix}

In order to evaluate analytically the atomic matrix element ${\cal M}%
_{at}^{II}$ in eq.(50) we use the expressions of the linear and quadratic
response given in Refs.\cite{f-m} and\cite{fhm}, respectively. The
exponential in the form factor formula (29) is written in the standard way
as a expansion in spherical harmonics. After integration over the angular
coordinates of the bound electron, one can write down the following
expression of the matrix element for two photon absorption/emission
\begin{eqnarray}
{\cal M}_{at}^{(II)} \left( \Omega^{\prime \pm}, \Omega^\pm \right) = -\frac{%
(\vec{s}\cdot\vec{q})^2}{8\pi^2 q^4} && \left[ _2{\cal {I}}_{10}^{21}({\tau}%
^{\prime \mp}, {\tau}^{\mp},q) + _2{\cal {I}}_{10}^{21}({\tau}^{\prime \pm},
{\tau}^{\pm},q) + _2{\cal {I}}_{101}({\tau}^{\mp},{\tau}^{\pm}, q) \right]
\nonumber \\
+ \frac{\vec{s}^{\;2}}{24 {\pi}^2 q^2} && \left[ _0{\cal {I}}_{10}^{01}({\tau%
}^{\mp}, {\tau}^{\mp},q) + _2{\cal {I}}_{10}^{21}({\tau}^{\mp}, {\tau}%
^{\mp},q)\right.  \nonumber \\
&& + _0{\cal {I}}_{10}^{01}({\tau}^{\pm}, {\tau}^{\pm},q) + _2{\cal {I}}%
_{10}^{21}({\tau}^{\pm}, {\tau}^{\pm},q)  \nonumber \\
&& + \left. _0{\cal {I}}_{101}({\tau}^{\mp},{\tau}^{\pm}, q)+ _2{\cal {I}}%
_{101}({\tau}^{\mp},{\tau}^{\pm}, q) \right] .  \label{tafa}
\end{eqnarray}
Here the upper signs correspond to absorption and the lower ones to emission
processes. If the polarization vector $\vec s$ is complex, its complex
conjugate should be taken in order to compute two photon emission. Using the
previous equation, one can write down in a straightforward manner the
general structure of ${\cal M}_{at}^{(II)}$ as
\begin{equation}
{\cal M}_{at}^{(II)}=\frac 1{2\pi ^2q^2}\left[ \frac{(\vec{s}\cdot \vec{q})^2%
}{q^2}{\cal T}_1 + {\vec s}^{\;2} {\cal T}_2\right] ,
\end{equation}
where ${\cal T}_1$ and ${\cal T}_2$ denote the following two combinations of
radial integrals
\begin{eqnarray}
{\cal T}_1&=& -\frac14 \left[ _2{\cal {I}}_{10}^{21}({\tau}^{\prime \mp}, {%
\tau}^{\mp},q) + _2{\cal {I}}_{10}^{21}({\tau}^{\prime \pm}, {\tau}^{\pm},q)
+ _2{\cal {I}}_{101}({\tau}^{\mp},{\tau}^{\pm}, q) \right] ,  \label{t1} \\
{\cal T}_2&=&\frac{1}{12} \left[ _0{\cal {I}}_{10}^{01}({\tau}^{\mp}, {\tau}%
^{\mp},q) + _2{\cal {I}}_{10}^{21}({\tau}^{\mp}, {\tau}^{\mp},q) + _0{\cal {I%
}}_{10}^{01}({\tau}^{\pm}, {\tau}^{\pm},q) + _2{\cal {I}}_{10}^{21}({\tau}%
^{\pm}, {\tau}^{\pm},q) \right.  \nonumber \\
& & \hspace{0.5cm}+ \left. _0{\cal {I}}_{101}({\tau}^{\mp},{\tau}^{\pm}, q)+
_2{\cal {I}}_{101}({\tau}^{\mp},{\tau}^{\pm}, q) \right] .  \label{t2a}
\end{eqnarray}
As a consequence, ${\cal T}_1$ and ${\cal T}_2$ depend on the momentum
transfer, $q$, and on the four parameters $\tau^{\prime \pm}$ and $%
\tau^{\pm} $.

The four radial integrals present in eq. (\ref{tafa}) are defined by
\begin{eqnarray}
_0{\cal {I}}_{101}(\Omega^{\prime},\Omega, q) & \equiv & \int dr r^2 {\cal B}%
_{101} (\Omega^{\prime}, r)\ j_{0}(qr)\ {\cal B}_{101}(\Omega, r) ,
\label{ir1} \\
_2{\cal {I}}_{101}(\Omega^{\prime},\Omega, q) & \equiv & \int dr r^2 {\cal B}%
_{101} (\Omega^{\prime}, r)\ j_{2}(qr)\ {\cal B}_{101}(\Omega, r) ,
\label{ir2} \\
_0{\cal {I}}_{10}^{01}(\Omega^{\prime}, \Omega,q) & \equiv & \int dr r^2
{\cal B }_{10}^{01}(\Omega^{\prime}, \Omega,r)j_0(qr)R_{10}(r) ,  \label{ir3}
\\
_2{\cal {I}}_{10}^{21}(\Omega^{\prime}, \Omega,q) &\equiv & \int dr r^2
{\cal B }_{10}^{21}(\Omega^{\prime}, \Omega,r)j_2(qr)R_{10}(r) ,  \label{ir4}
\end{eqnarray}
where $j_l(qr)$ are spherical Bessel functions of order the $l$. $R_{10}$ is
the radial function of the hydrogen ground state, ${\cal B}_{101}$ is
defined by eq.(32) of Ref.\cite{f-m}, and ${\cal B}_{10}^{01}$, ${\cal B}%
_{10}^{21}$ by eqs.(43)-(44) of Ref.\cite{fhm}.

We have obtained analytic expressions for the radial integrals in eqs.(\ref
{ir1})-(\ref{ir4}). The first two integrals, $_0{\cal {I}}_{101}$ and $_2%
{\cal {I}}_{101}$, are related to the linear response. They may be
expressed, in terms of Appell functions all of which depend on the same
variables, namely
\begin{eqnarray}
\xi_1 &=& \frac{1-\tau^{\prime}}{2} ,\quad \zeta_1 = \frac{%
(1-\tau^{\prime})\tau}{\tau+\tau^{\prime}-iq\tau\tau^{\prime}} .
\label{varf1}
\end{eqnarray}
Note that, given the definitions (\ref{ir1})-(\ref{ir2}), these two radial
integrals are symmetric with respect to the parameters $\tau$ and $%
\tau^{\prime}$. We present below the expression for the first integral
\begin{eqnarray}
& & _2{\cal {I}}_{101}(\tau^{\prime},\tau, q) =  \label{int1} \\
& &\hspace{1.cm} \frac{8 \tau\tau^{\prime}}{q (2-\tau^{\prime})(1+\tau)}
\left[ \frac{2}{1+\tau^{\prime}}\right]^{2+\tau^{\prime}} \sum_{m=0}^{\infty}%
{a_m^{\prime}}(\tau) \left[\frac{1-\tau}{\tau}\right]^m \sum_{p=0}^{2} \frac{%
(p+1)(p+2)}{(2q)^p }\frac{(3+m-p)!}{(2-p)!}  \nonumber \\
& & \hspace{2cm}\times Re\left\{i^{p-3} \chi_1 ^{4+m-p}
F_1(2-\tau^{\prime},-1-\tau^{\prime},4+m-p,3-\tau^{\prime};\xi_1,\zeta_1)%
\right\} .  \nonumber
\end{eqnarray}
and that for the combination $_0{\cal {I}}_{101}({\tau}^{\prime},{\tau}, q)+
_2{\cal {I}}_{101}({\tau}^{\prime},{\tau}, q)$
\begin{eqnarray}
& & _0{\cal {I}}_{101}({\tau}^{\prime},{\tau}, q)+ _2{\cal {I}}_{101}({\tau}%
^{\prime},{\tau}, q) =  \label{int2} \\
& & \hspace{.75cm} -\frac{24 \tau\tau^{\prime}}{q^2(2-\tau^{\prime})(1+\tau)}
\left[\frac{2}{1+\tau^{\prime}}\right]^{2+\tau^{\prime}} \sum_{m=0}^{\infty}
{a_m^{\prime}}(\tau) \left[\frac{1-\tau}{\tau}\right]^m (m+1)!  \nonumber \\
& & \hspace{1.75cm}\times Re\left\{ \sum_{p=1}^{2}\left(\frac{i}{q}%
\right)^{2-p}(m+p)^{p-1} \chi_1^{1+m+p}
F_1(2-\tau^{\prime},-1-\tau^{\prime},1+m+p,3-\tau^{\prime}, \xi_1,\zeta_1)
\right\} .  \nonumber
\end{eqnarray}
Here the following notations have been used
\begin{eqnarray}
\chi_1 &=& \frac{\tau\tau^{\prime}}{\tau+\tau^{\prime}-iq\tau\tau^{\prime}},
\\
{a_m^{\prime}}(\tau) & = & \frac{1}{m!} \frac{1}{2-\tau+m} \;
_2F_1\left(1,-1-\tau,3-\tau+m ,\frac{\tau-1}{\tau+1}\right),  \label{am}
\end{eqnarray}
where $_2F_1$ denotes the Gauss function. We stress that all the definitions
and notations used in this paper for the hypergeometric functions are those
given in Ref.\cite{hyp}. To avoid possible confusions, we remind that
according to eq.(\ref{tafa}) the parameters $\tau$ and $\tau^{\prime}$ take
the values $\tau^{\pm}$ given by eq.(\ref{tau1}).

The other two radial integrals are related to the quadratic response; they
are expressed as series of hypergeometric functions $F_2$ instead of Appell
functions $F_1$. Although all the functions $F_2$, which are involved,
depend also on the same variables, namely
\begin{equation}
\xi_2 = \frac{\tau-\tau^{\prime}}{2\tau} ,\quad \zeta_2 = \frac{%
\tau+\tau^{\prime}}{\tau(1-\tau^{\prime}+iq\tau^{\prime})} ,
\end{equation}
the integrals $_0{\cal {I}}_{10}^{01}$ and $_2{\cal {I}}_{10}^{21}$ are no
more symmetric with respect to the parameters $\tau$ and $\tau^{\prime}$,
that take the values in eqs.(\ref{tau1}) and (\ref{tau2}), respectively. The
expressions for the integral $_2{\cal {I}}_{10}^{21}(\tau^{\prime},\tau,q)$
as well as that for the combination $_0{\cal {I}}_{10}^{01}(\tau^{\prime},%
\tau,q) +_2{\cal {I}}_{10}^{21}(\tau^{\prime},\tau,q)$ are written down
below
\begin{eqnarray}
& & _2{\cal {I}}_{10}^{21}(\tau^{\prime},\tau,q) = \frac{2}{q}(1+\tau) \left[%
\frac{\tau^{\prime}(\tau+\tau^{\prime})}{\tau(\tau^{\prime}+ 1 )}\right]^5
\sum_{m=0}^{\infty}d_m^{\prime}(\tau^{\prime},\tau) \left[\frac{%
\tau^{\prime}(\tau-1)}{\tau(\tau^{\prime}+1)}\right]^m \frac{(6)_m}{%
3-\tau^{\prime}+m}  \label{int3} \\
& &\hspace{2.cm}\times \sum_{p=0}^{2}\frac{(p+1)(p+2)(3-p)}{%
(2q\tau^{\prime})^{p}} Re \left\{ i^{p-1}\chi_2^{4-p}
F_2(6+m,1,4-p,4-\tau^{\prime}+m,6;\xi_2,\zeta_2)\right\}  \nonumber
\end{eqnarray}
and
\begin{eqnarray}
_0{\cal {I}}_{10}^{01}(\tau^{\prime},\tau,q) &+&_2{\cal {I}}%
_{10}^{21}(\tau^{\prime},\tau,q)  \label{int4} \\
&= & \frac{24\tau\tau^{\prime}}{q\left(1+\tau\right)} \left[\frac{%
\tau^{\prime}(\tau+\tau^{\prime})}{\tau(1+\tau^{\prime})}\right]^3
\sum_{m=0}^{\infty} {c_m^{\prime}}(\tau^{\prime},\tau) \frac{(5)_m}{%
2-\tau^{\prime}+m} {\left[\frac{\tau^{\prime}(\tau-1)}{\tau(\tau^{\prime}+1)}%
\right] }^{m}  \nonumber \\
& & \hspace{1.cm}\times Re\left\{i\chi_2^{3}
F_2(5+m,1,3,3-\tau^{\prime}+m,5;\xi_2,\zeta_2)\right\}  \nonumber \\
&+&\frac{6\left(1+\tau \right)}{q^2\tau^{\prime}} \left[\frac{%
\tau^{\prime}(\tau+\tau^{\prime})}{\tau(\tau^{\prime}+ 1 )}\right]^5
\sum_{m=0}^{\infty}d_m^{\prime}(\tau^{\prime},\tau) \frac{(6)_m}{%
3-\tau^{\prime}+m} \left[\frac{\tau^{\prime}(\tau-1)}{\tau(\tau^{\prime}+1)}%
\right]^m  \nonumber \\
& &\hspace{1.cm} \times Re\left\{\sum_{p=1}^{2} \left(\frac{i}{%
2q\tau^{\prime}}\right)^{p-2} \chi_2^{1+p}
F_2(6+m,1,1+p,4-\tau^{\prime}+m,6;\xi_2,\zeta_2)\right\}  \nonumber \\
&-& \frac{12\tau \tau^{\prime}}{q \left(1+\tau\right)} \left[\frac{%
\tau^{\prime}(\tau+\tau^{\prime})}{\tau(1+\tau^{\prime})}\right]^3
\sum_{m=0}^{\infty} {b_m^{\prime}}(\tau^{\prime},\tau) {\left[\frac{%
\tau^{\prime}(\tau-1)}{\tau(\tau^{\prime}+1)}\right] }^m  \nonumber \\
& & \hspace{1.cm}\times Re\left\{i \chi_2^{3} \left[ 2 \frac{%
\tau+\tau^{\prime}}{\tau}\frac{(5)_m }{3-\tau^{\prime}+m} F_2(5+m,1,3,4-%
\tau^{\prime}+m,5;\xi_2,\zeta_2)\right.\right.  \nonumber \\
& & \hspace*{3.cm} - \chi_2^{-1}\frac{(4)_m}{3-\tau^{\prime}+m}
F_2(4+m,1,2,4-\tau^{\prime}+m,4;\xi_2,\zeta_2)  \nonumber \\
& & \hspace*{3.cm} + \left.\left. \chi_2^{-1} \frac{(4)_m}{1-\tau^{\prime}+m}
F_2(4+m,1,2,2-\tau^{\prime}+m,4;\xi_2,\zeta_2)\right]\right\} ,  \nonumber
\end{eqnarray}
where $(n)_m$ denotes the Pochhammer's symbol and the following notations
have been use
\begin{equation}
\chi_2 = \frac{1}{\tau^{\prime}-1-iq\tau^{\prime}}
\end{equation}
and
\begin{eqnarray}
b_m^{\prime}(\tau^{\prime},\tau) & = &\frac{1}{m!} \frac{1}{2-\tau+m} \;
_2F_1\left(1,-1-\tau,3-\tau+m , \delta_2\right)  \label{bm} \\
c_m^{\prime}(\tau^{\prime},\tau) & = &\frac{1}{m!} \left[2\frac{%
\tau^{\prime}(\tau-1)}{\tau(\tau^{\prime}+1)} \frac{1}{3-\tau+m} \;
_2F_1\left(1,-1-\tau,4-\tau+m , \delta_2\right) \right.  \nonumber \\
&& \hspace{1.1cm} - \left.\frac{\tau-\tau^{\prime}}{\tau}\frac{1}{2-\tau+m}%
\; _2F_1\left(1,-1-\tau,3-\tau+m , \delta_2\right) \right]  \label{cm} \\
d_m^{\prime}(\tau^{\prime},\tau) & = & \frac{1}{m!} \left[ - \left(\frac{%
\tau-1}{\tau+1}\right)^2 \frac{1}{4-\tau+m}\; _2F_1\left(1,-1-\tau,5-\tau+m
, \delta_2\right)\right.  \nonumber \\
&&\hspace{2.5cm}+\left.\frac{1}{2-\tau+m} \;
_2F_1\left(1,-3-\tau,3-\tau+m,\delta_2\right) \right]  \label{dm}
\end{eqnarray}
with
\begin{equation}
\delta_2=\frac{(\tau-1)(\tau-\tau^{\prime})}{(\tau+1)(\tau+\tau^{\prime})}.
\end{equation}
The expressions of the radial integrals in eqs.(\ref{int1})-(\ref{int2}) and
(\ref{int3})-(\ref{int4}) were written down for real values of the
parameters $\tau ^{\pm}$ and $\tau ^{\prime \pm}$.

The {\it one} photon resonances discussed in section IV are related to the
poles of the four radial integrals (\ref{ir1})-(\ref{ir4}). They occur for $%
\tau ^{+}=n$, which corresponds to $\omega =|{\rm E}_1|\left(
1-n^{-2}\right) $ with $n\geq 2$. {\it Two} photon resonances are related to
poles of $_0{\cal I}_{10}^{01}$ and $_2{\cal I}_{21}^{01}$, only. They occur
for $\tau ^{\prime +}=n$ and they correspond to $2\omega =|{\rm E}_1|\left(
1-n^{-2}\right) $. The integral $_0{\cal I}_{10}^{01}$ has poles for any
value of $n$, while we should note that $_2{\cal I}_{21}^{01}$ has poles
only for $\tau ^{\prime +}\geq 3$. This explains the absence of a resonance
at $\omega =5.1$ eV in the frequency dependence of the nonlinear signal for $%
N=2$ if the laser field is circularly polarized, see Figure 3(a), because
only $_2{\cal I}_{21}^{01}$ enters the expression for ${\cal T}_1$.\medskip
\newline
{\Large Acknowledgments}

This work has been supported by the Jubilee Foundation of the Austrian
National Bank under project number 6211.

\newpage

FIGURE CAPTIONS

Fig.1: $d\sigma_N/d\Omega$ are presented as a function of the scattering
angle, $\theta$, for a laser field ${\cal E}_0$=10$^8$ V/cm for the geometry
{\bf CP3} (full line) and {\bf LP3} (dotted line). The initial projectile
energy is $E_i$=100 eV. Panels $a_1$, $b_1$, and $c_1$ on the left hand side
refer to $\omega$=5 eV and the right hand panels correspond to $\omega$=2
eV. In panels $a_1$ and $a_2$ are also presented the field-free differential
cross sections (dashed lines).

Fig.2: $d\sigma_1/d\Omega$ normalized with respect to the field intensity,
is shown as a function of the laser frequency for the geometry {\bf CP3} at
the scattering angle $\theta=5^{\circ}$ and initial projectile energy $E_i$%
=100 eV. Also plotted are the corresponding data of the Bunkin-Fedorov
formula (dotted line) and the atomic contribution of the first order
dressing (dotted-dashed line).

Fig.3(a): $d\sigma_2/d\Omega$ normalized with respect to $I^2$, is shown as
a function of $\omega$ for the geometry {\bf CP3} at the scattering angle $%
\theta=5^{\circ}$ and initial projectile energy $E_i$=100 eV. Also plotted
are the results of the Bunkin-Fedorov formula (dotted line) and the atomic
contribution due to first order (dotted-dashed line) and second order
dressing (long dashed line). (b) same as in Fig.3(a), but for the geometry
{\bf LP3}.

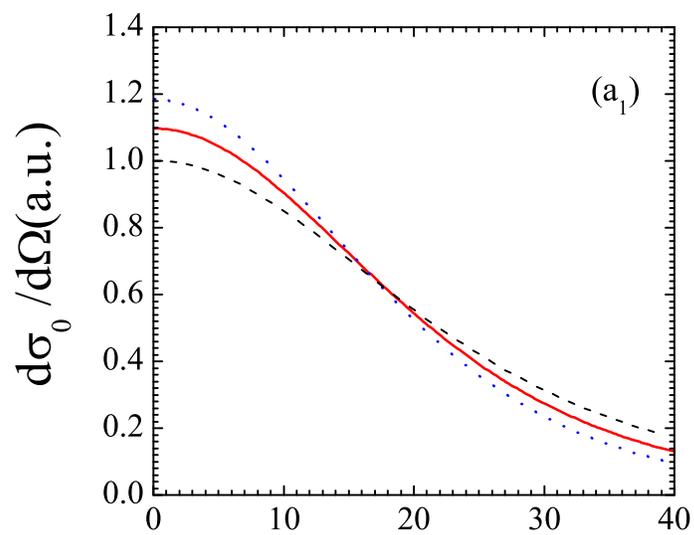
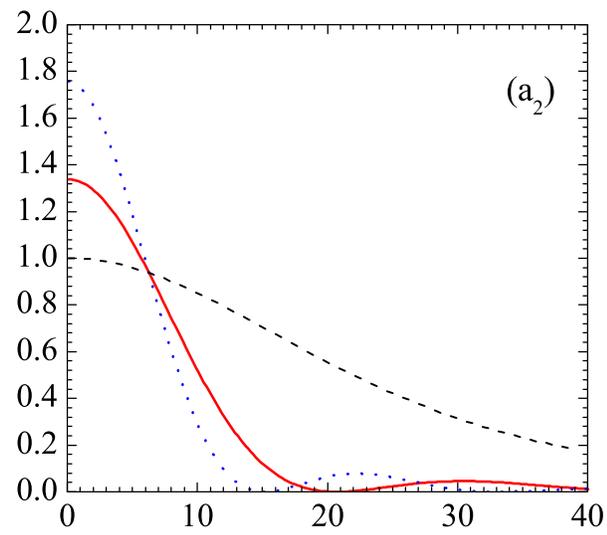
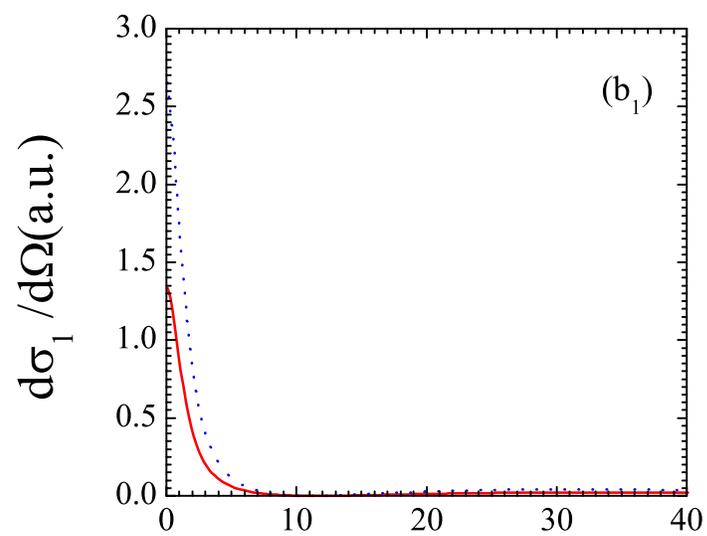
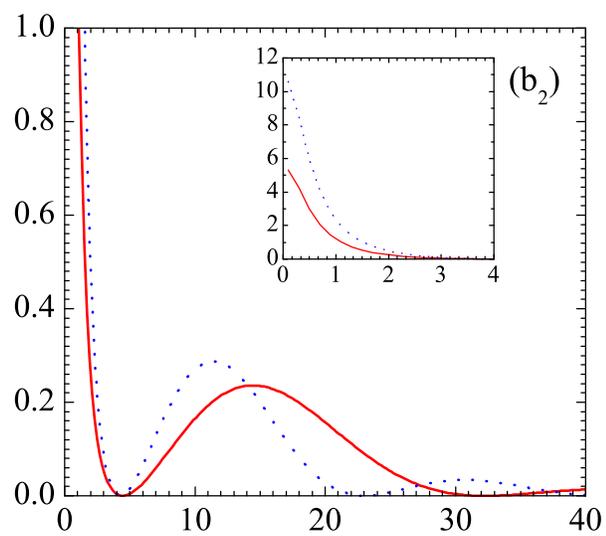
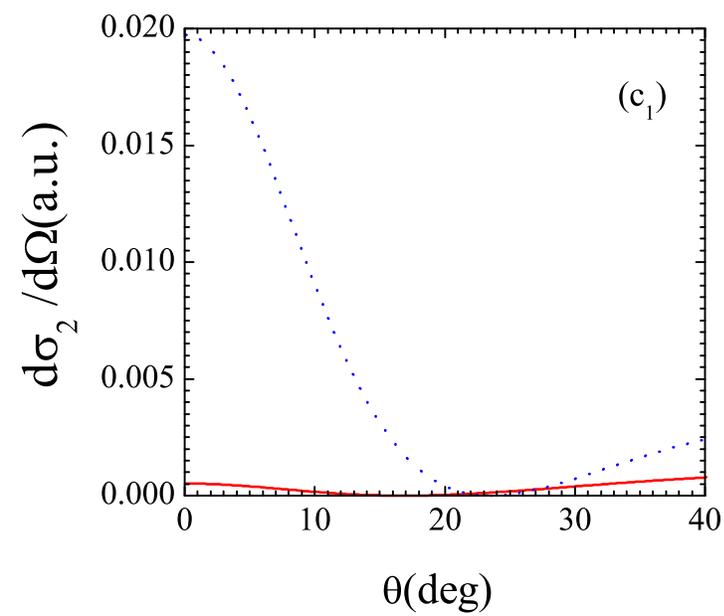
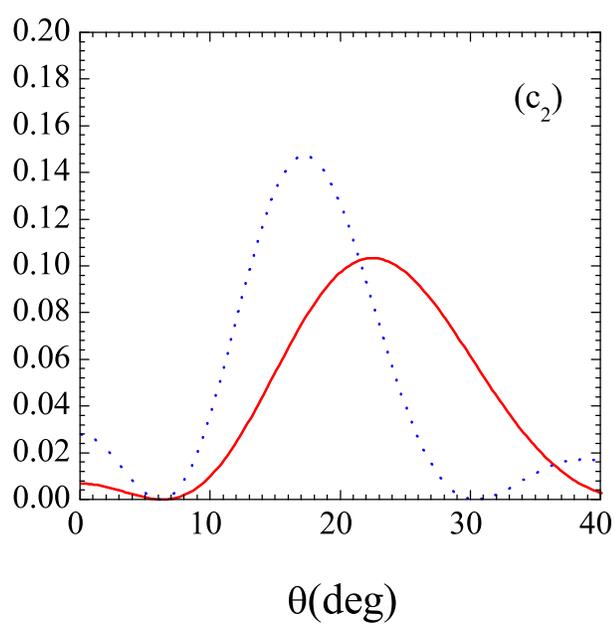
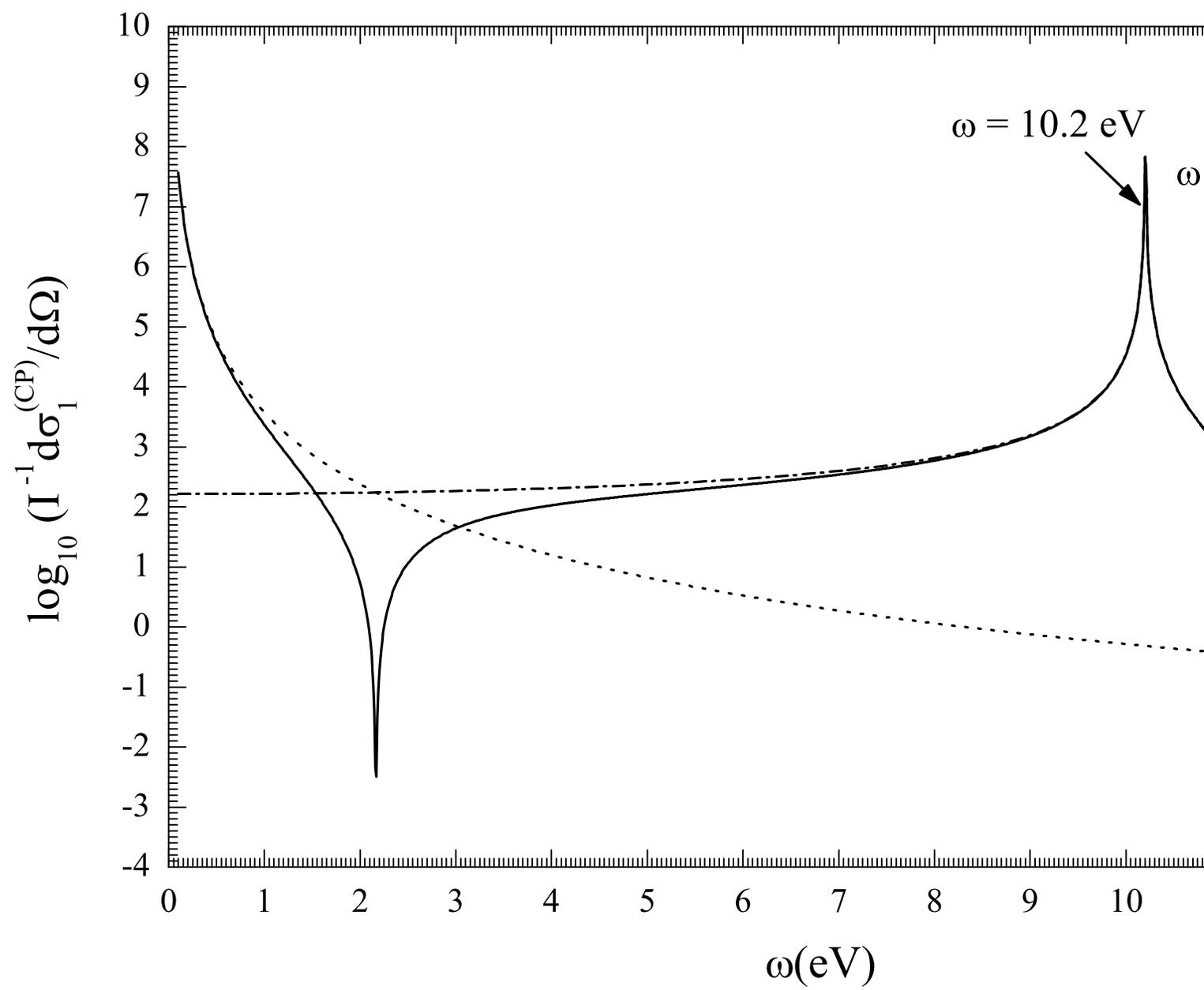
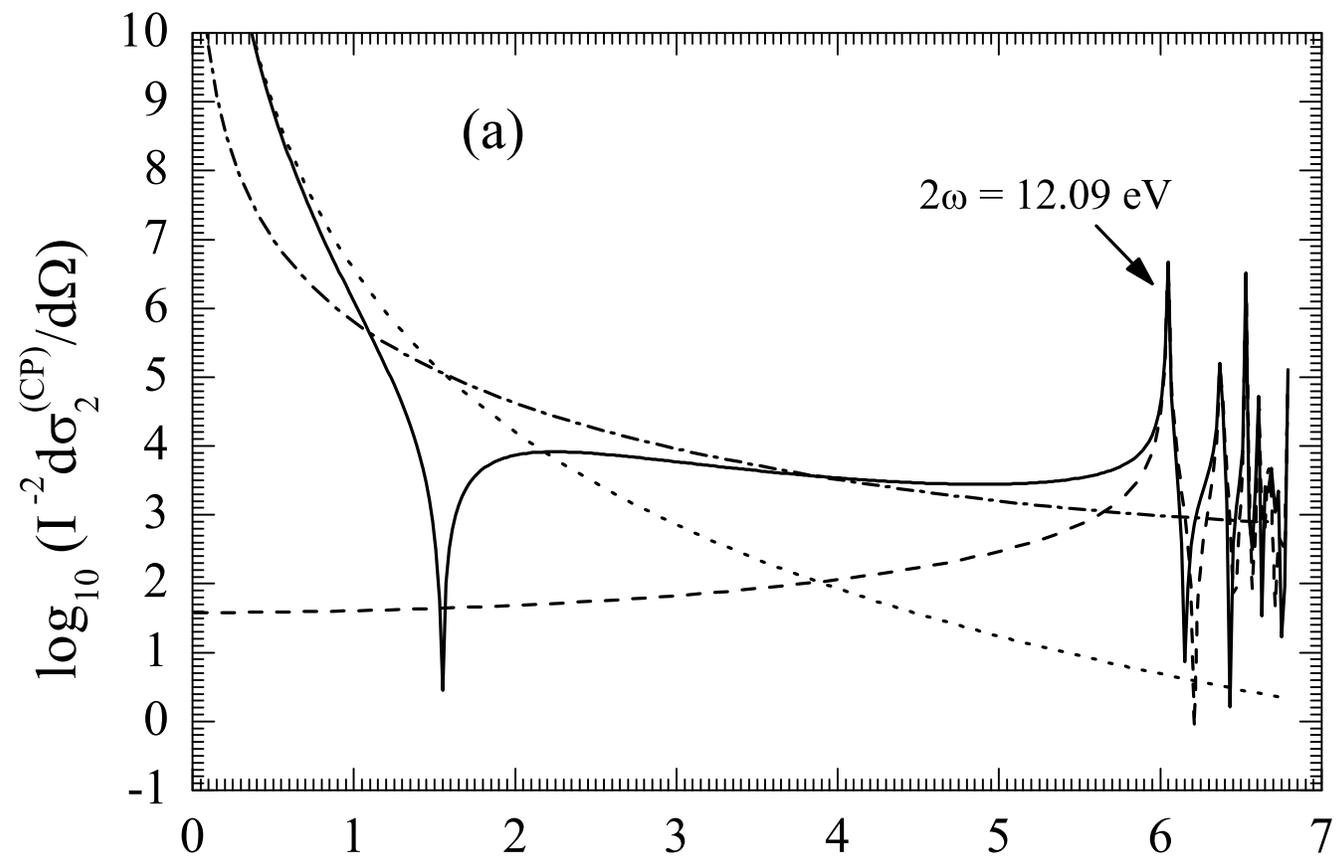
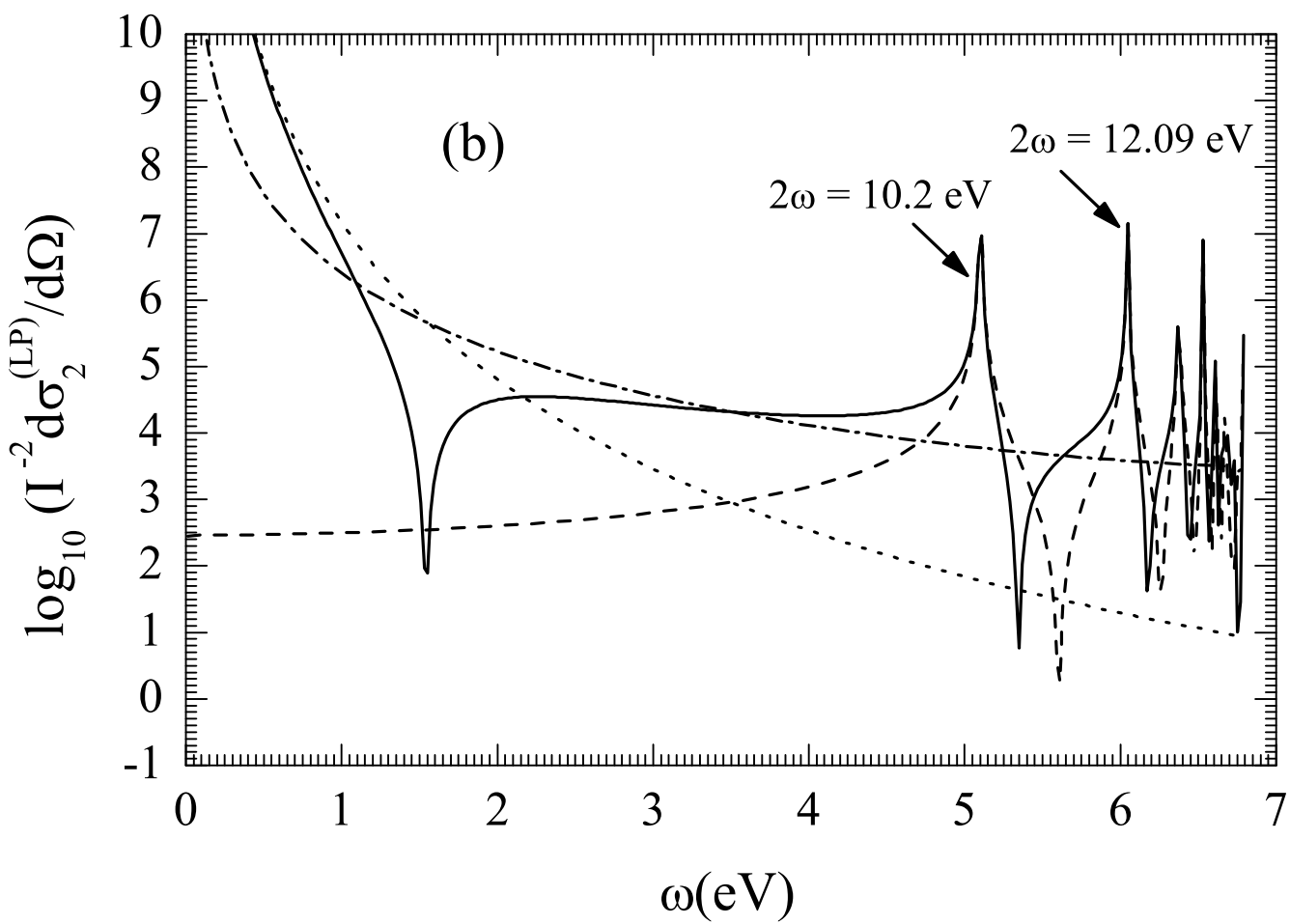
\end{document}